\documentclass{emulateapj}

\usepackage{epsfig}
\usepackage{graphicx}
\def\gta{\mathrel{\hbox{\rlap{\hbox{\lower4pt\hbox{$\sim$}}}\hbox{$>$}}}}

\shorttitle{RCS}
\shortauthors{Krick et al.}

\begin{document}
\newcommand\msun{\hbox{M$_{\odot}$}}
\newcommand\lsun{\hbox{L$_{\odot}$}}
\newcommand\magarc{mag arcsec$^{-2}$}
\newcommand\h{$h_{70}^{-1}$}

\title{\bf Galaxy Clusters in the IRAC Dark Field I: \\ Growth of the red sequence}

\author{J.E.~Krick \altaffilmark{1}, J.A.~Surace \altaffilmark{1}, D.~Thompson \altaffilmark{2}, M.L.N.~Ashby \altaffilmark{3}, J.L.~Hora \altaffilmark{3}, V.~Gorjian \altaffilmark{4}, and L.~Yan \altaffilmark{1}}

\altaffiltext{1}{Spitzer Science Center, MS 220--6,
California Institute of Technology, Jet Propulsion Laboratory,
Pasadena, CA 91125, USA}
\altaffiltext {2}{Large Binocular Telescope Observatory, University of Arizona, Tucson, AZ 85721, USA}
\altaffiltext {3}{Harvard-Smithsonian Center for Astrophysics, 60 Garden Street, Cambridge MA 02138, USA}
\altaffiltext {4}{Jet Propulsion Laboratory, California Institute of Technology, Pasadena, CA, 91109, USA}
\email{jkrick@caltech.edu}

\begin{abstract} 

  Using three newly identified galaxy clusters at $z\sim1$
  (photometric redshift) we measure the evolution of the galaxies
  within clusters from high redshift to the present day by studying
  the growth of the red cluster sequence.  The clusters are located in
  the Spitzer Infrared Array Camera (IRAC) Dark Field, an extremely
  deep mid-infrared survey near the north ecliptic pole with
  photometry in 18 total bands from X-ray through far-IR. Two of the
  candidate clusters are additionally detected as extended emission in
  matching Chandra data in the survey area allowing us to measure
  their masses to be $M_{500}= 6.2 \pm 1.0 \times 10^{13}$ and $3.6
  \pm 1.1 \times 10^{13}$ \msun.  For all three clusters we create a
  composite color magnitude diagram in rest-frame B-K using our deep
  HST and Spitzer imaging.  By comparing the fraction of low
  luminosity member galaxies on the composite red sequence with the
  corresponding population in local clusters at $z=0.1$ taken from the
  COSMOS survey, we examine the effect of a galaxy's mass on its
  evolution.  We find a deficit of faint galaxies on the red sequence
  in our $z\sim1$ clusters which implies that more massive galaxies
  have evolved in clusters faster than less massive galaxies, and that
  the less massive galaxies are still forming stars in clusters such
  that they have not yet settled onto the red sequence.

\end{abstract}


\section{Introduction} 

The redshift range from $z=1$ to the present day is a particularly
dynamic epoch in the history of groups and clusters as evidenced by
the evolution of the morphology-density relation and increasing
fraction of blue galaxies with increasing redshift \citep{capak2007,
  butcher1984}. Interestingly, cluster ellipticals at $z\sim1$ already
have a narrow distribution of red colors \citep[the red cluster
sequence (RCS);][]{blakeslee2003,vandokkum2001}.  There is some debate
about the mechanism by which these cluster galaxies arrive onto the
red sequence.  It is difficult to distinguish whether these red
ellipticals all formed their stars and did their merging at $z>3$,
then stopped forming stars when they entered the cluster environment
\citep{ford2004}; or if they are the product of the merging of
gas-poor systems which do not produce star formation
\citep{vandokkum2005}.

We investigate whether the red population is still in the process of
forming at $z=1$ or if indeed assembly has already finished at higher
redshift by studying the presence of the faint end of the RCS at $z=1$
and comparing it to the present epoch.  We measure the ratio of faint
to bright RCS galaxies in a sample of three $z\sim1$ clusters from the
IRAC Dark Field (described below).  These clusters have the benefit of
extremely deep $3.6\,\micron$ data which allows us to study the faint
end of the luminosity function at rest frame near-IR, which traces the
peak of the spectral energy distribution in galaxies.  A confirmed
deficit of faint galaxies on the RCS would imply that more massive
galaxies have evolved in clusters faster than less massive galaxies.
A constant fraction of faint red galaxies between $z=1$ and the
present would require a formation mechanism where galaxies of all
masses have already joined the red sequence at redshifts higher than
one.  There is evidence that the faint end of the RCS is not
completely in place by z=0.8 \citep[][and references
therein]{delucia2007,koyama2007}, although at least some clusters at
these redshifts appear to have complete RCSs to M*+3.5
\citep{andreon2006}.

These questions are ideally addressed with deep infrared surveys of
clusters at high redshift.  In the last four years deep and wide area
surveys in the mid-infrared using the {\it Spitzer Space Telescope}
have substantially opened a new window on galaxy and star formation at
$0<z<3$.  {\it Spitzer} now routinely produces imaging of
large fields to higher resolution and fainter depths than previously
possible.  IRAC, the mid-infrared camera on-board Spitzer, takes
images at 3.6, 4.5, 5.8, and 8.0\,\micron \citep{fazio2004}.  The
shorter wavelengths provide a direct measurement of the stellar
content of galaxies at redshifts as high as three.  The longer
wavelength channels sample emission from polycyclic aromatic
hydrocarbons (PAHs) in low redshift galaxies, as well as direct
thermal emission from hot dust.  The deepest such survey is the dark
current calibration field for the mid-infrared camera, commonly known
as the ``IRAC Dark Field''.  These deep mid-IR data are supplemented
by additional 14 band photometry including Palomar u',g',r',i', {\it
  HST/ACS} F814W, MMT $z^\prime$, Palomar J, H,\& K , Akari 11 \&
$15\,\micron$, Spitzer MIPS 24 \& $70\,\micron$, and Chandra ACIS-I
imaging.  This extensive, multiwavelength dataset gives us the distinctly unique
opportunity to study the development of the red sequence in galaxy
clusters at redshift one.  The second paper on galaxy clusters in this
series will discuss the role of star formation in the evolution of
clusters at $z=1$ by examining the {\it Spitzer} 24\,\micron \ data in
conjunction with the morphological information from the HST ACS
dataset.

This paper is structured in the following manner. In
\S\ref{observations} \& \S\ref{photz} we discuss the data and derived
photometric redshift determination.  Details of the cluster search and
cluster properties are presented in \S\ref{search}.  In \S\ref{cmds}
\& \S\ref{results} we present the color magnitude diagrams of the
candidate clusters and results of the faint-to bright ratios of red
sequence galaxies.  In \S\ref{discuss} we discuss the implications for
the evolution of cluster galaxies.  Throughout this paper we use
$H_0=70$km/s/Mpc, $\Omega_M$ = 0.3, $\Omega_\Lambda$ = 0.7.  With this
cosmology, the luminosity distance at z=1 is 6607 Mpc, but the angular
diameter distance is a factor of $(1+z)^2$ less, or 1652 Mpc.  All
photometry is quoted in the AB magnitude system.

%%%%%%%%%%%%%%%%%%%%%%%%%%%%%%%%%%%%%%%%%%%%%%%%%%%%%%5

\section{Observations \& Data Reduction}
\label{observations}

\subsection{The IRAC Dark Field}

The survey region is the IRAC Dark Field, centered at approximately
17h40m +69d.  The field is located a few degrees from the north
ecliptic pole (NEP) in a region which is darker than the actual pole
and is in the Spitzer continuous viewing zone so that it can be
observed any time IRAC is powered on for observing.  These observing
periods are called instrument ``campaigns'', and occur roughly once
every three to four weeks and last for about a week. Sets of long
exposure frames are taken on the Dark Field at least twice during each
campaign totaling roughly four hours of integration time per campaign,
and these data are used to derive dark current/bias frames for each
channel.  The dark frames are used by the pipeline in a manner similar
to ``median sky'' calibrations as taken in ground-based near-infrared
observing to produce the Basic Calibrated Data (BCD) for all science
observations. Each set of dark calibration observations collects
roughly two hours of integration time at the longest exposure times in
each channel.

The resulting observations are unique in several ways.  The Dark Field
lies near the lowest possible region of zodiacal background, the
primary contributor to the infrared background at these wavelengths,
and as such is in the region where the greatest sensitivity can be
achieved in the least amount of time.  The area was also chosen
specifically to be free of bright stars and very extended galaxies,
which allows clean imaging to very great depth. The observations are
done at many position angles (which are a function of time of
observation) leading to a more uniform final {\sc psf}.  Finally,
because the calibration data are taken directly after anneals, they
are more free of artifacts than ordinary guest observer (GO)
observations.  Over the course of the mission, the observations have
filled in roughly uniformly a region 20\arcmin \ in diameter.  This
has created the deepest ever mid-IR survey, exceeding the depth of the
deepest planned regular Spitzer surveys over several times their area.
Furthermore, this is the only field for which a 5+year baseline of
mid-IR periodic observations is expected.

The IRAC data are complemented by imaging data in 14 other bands with
facilities including Palomar, MMT, HST, Akari, Spitzer MIPS, and
Chandra ACIS-I.  Although the entire dark field is $> 20\arcmin$ in
diameter, because of spacecraft dynamics the central $\sim 15\arcmin$
is significantly deeper and freer of artifacts.  Therefore, it is this
area which we have matched with the additional observations.  The
entire dataset will be presented in detail in a future paper (Krick et
al, in prep).  For completeness we briefly discuss here the Spitzer
IRAC, HST ACS, and Chandra ACIS datasets as they are the most critical
to this work.  All space-based datasets are publicly available through
their respective archives.

\subsection{Spitzer/IRAC}
\label{irac}

This work is based on a preliminary combination of 75 hours of IRAC
imaging, which is $\approx$30\% of the expected depth not including a
possible warm mission.  Even these 75 hours go well into the confusion
limit of IRAC and so the additional exposure time will not add a lot
of sensitivity.  The Basic Calibrated Data (detector image level)
product produced by the Spitzer Science Center was further reduced
using a modified version of the pipeline developed for the SWIRE
survey \citep{surace2005}.  This pipeline primarily corrects image
artifacts and forces the images onto a constant background
(necessitated by the continuously changing zodiacal background as seen
from Spitzer). The data were coadded onto a regularized 0.6\arcsec \
grid using the MOPEX software developed by the Spitzer Science Center.

Experiments with DAOPHOT demonstrate that nearly all extragalactic
sources are marginally resolved by IRAC, particularly at the shorter
wavelengths, and hence point source fitting is inappropriate.
Instead, photometry is done using the high spatial resolution ACS data
as priors for determining the appropriate aperture shape for
extracting the Spitzer data.  We do this by first running source
detection and photometric extraction on the coadded IRAC images using
a matched filter algorithm with image backgrounds determined using the
mesh background estimator in SExtractor (Bertin et al. 1995) .  This
catalog is merged with the HST ACS catalog.  For every object in that
catalog ,if the object is detected in ACS then we use the ACS shape
parameters to determine the elliptical aperture size for the IRAC
images.  ACS shape parameters are determined by SExtractor on
isophotal object profiles after deblending, such that each ACS pixel
can only be assigned to one object (or the background).  For objects
which are not detected in ACS, but which are detected in IRAC, we
simply use the original IRAC SExtractor photometry.  Because of the
larger IRAC beam, we impose a minimum semi-major axis radius of
2\arcsec.  In all cases aperture corrections are computed individually
from PSF's provided by the Spitzer Science Center based on the aperture
sizes and shapes used for photometry.  

Final aperture photometry was performed using custom extraction
software written in IDL and based on the APER and MASK$\_$ELLIPSE
routines with the shape information from SExtractor, from either ACS
or IRAC as described above, using local backgrounds.  Because we use
local backgrounds, the measured fluxes of objects near the confusion
limit should have a larger scatter than those non-confused objects,
but will on average be the correct flux.  This will have the effect of
adding scatter to the CMD described below, but will not cause trends
of movement for the faint objects in color space.  Additionally, this
will not effect the photometric redshifts, as it will likely shift all
IRAC points up or down, but not relative to each other.  The final
$95\%$ completeness limit at 3.6\,\micron\ is 0.2\,$\mu$Jy or 25.7 AB
magnitude as calculated from a number count diagram by measuring where
the number of observed objects drops below 95\% of the expected number
of objects, where the expected numbers are calculated by fitting a
straight line to the brighter flux number counts.

\subsection{HST/ACS}
\label{acs}
The {\it HST} observations consist of 50 orbits with the ACS
comprising 25 separate pointings, all with the F814W filter (observed
I-band).  Within each pointing eight dithered images were taken for
cosmic ray rejection and to cover the gap between the two ACS CCDs.
The ACS pipeline {\it calacs} was used for basic reduction of the
images.  Special attention was paid to bias subtraction, image
registration, and mosaicing.  Pipeline bias subtraction was
insufficient because it does not measure the bias level individually
from each of the four amplifiers used by ACS.  We make this correction
ourselves by subtracting the mean value of the best fit Gaussian to
the background distribution in each quadrant.  Due to distortions in
the images, registration and mosaicing was performed with a
combination of IRAF's {\it tweakshifts, multidrizzle,} and {\it SWarp
  v.2.16.0} from Terapix.  The actual task of mosaicing the final
image was complicated by the large image sizes.  The single combined
mosaic image is 1.7GB and reading in all 200 images (160Mb each) for
combination is impossible for most software packages.

The final combined ACS image is $\sim 15\arcmin$ diameter coincident
with the deepest part of the IRAC Dark Field and is made with the
native $0.05\arcsec$ per pixel resolution.  Photometry was performed
in a standard manner with SExtractor. The $3\sigma$ detection limit
for point sources is $I = 28.6(AB)$.  The area common to both IRAC and
ACS contains $\sim 51,000$ detected sources.

All cluster galaxies are detected in the mid-infrared data, and the ACS
data are used as priors for extraction of the mid-IR photometry.
Because the cluster galaxies are detected in the optical ACS images, and
cluster membership is derived specifically based on the optical data,
it is the completeness of the optical data thats sets the fundamental
detection limits. Thus our ability to examine the faint end of the
cluster (rest frame) K-band luminosity function is ultimately limited
by the optical data, not the infrared data.

\subsection{Chandra/ACIS-I}
\label{chandra}

The field-of-view of the Chandra ACIS-I is 17\arcmin \ (for the
central four chips) , well-matched to the deepest central area of the
Dark Field. As a result, only a single pointing was required.  The
100ks observation was broken into three separate observations at
different pointing angles to reduce the effect of the gap between
chips.  The data was reduced using the newest version of the Chandra
Interactive Analysis of Observations software (CIAO 4.0).  Specific
attention was paid to destreaking, bad pixels, and background flares.
The task {\it merge\_all} was then used to combine event files from
the three observations into a final event file and image.  The final
combined Chandra image has $0.5\arcsec$ per pixel resolution. Blind
pointing is expected to be 1\arcsec.  Aperture photometry for extended
sources is done manually (see \S\ref{xray}).
%---------------------------------------------------------------------------------

\section{Photometric redshifts}
\label{photz}

The combined IRAC and ACS catalog contains over $50,000$ objects which
makes acquisition of spectroscopic redshifts impractical.  Even
confirmation spectroscopy of red galaxies at $z=1$ in our three
candidate clusters will require multiple nights on 8-10m class telescopes.
In lieu of spectroscopy we use our extensive multi-wavelength,
broad-band catalog to build spectral energy distributions (SEDs) and
derive photometric redshifts.  These SEDs are fit with template
spectra derived from galaxies in the Spitzer wide area infrared survey
survey \citep[SWIRE;][]{Polletta2007}.  Since the SWIRE templates are
based on Spitzer observations we find them the best choice to use as
models for this dataset.  Photometric redshifts are calculated using
{\it Hyperz}; a chi-squared minimization fitting program including a
correction for Galactic reddening \citep{bolzonella2000,
  calzetti2000}.  We emphasize that in this work photometric redshifts
are used only as a first step to find cluster candidates, and not to
determine membership within the clusters.

%---------------------------------------------------------------------------------
\section{Cluster Search}
\label{search}

The mid-IR wavelengths of IRAC are well suited to find galaxies at
redshift one because the stellar peak of the spectral
energy distribution has red-shifted into those bands.  We exploit this
fact in a search for clusters.

\begin{figure}
\epsscale{0.3}
\plotone{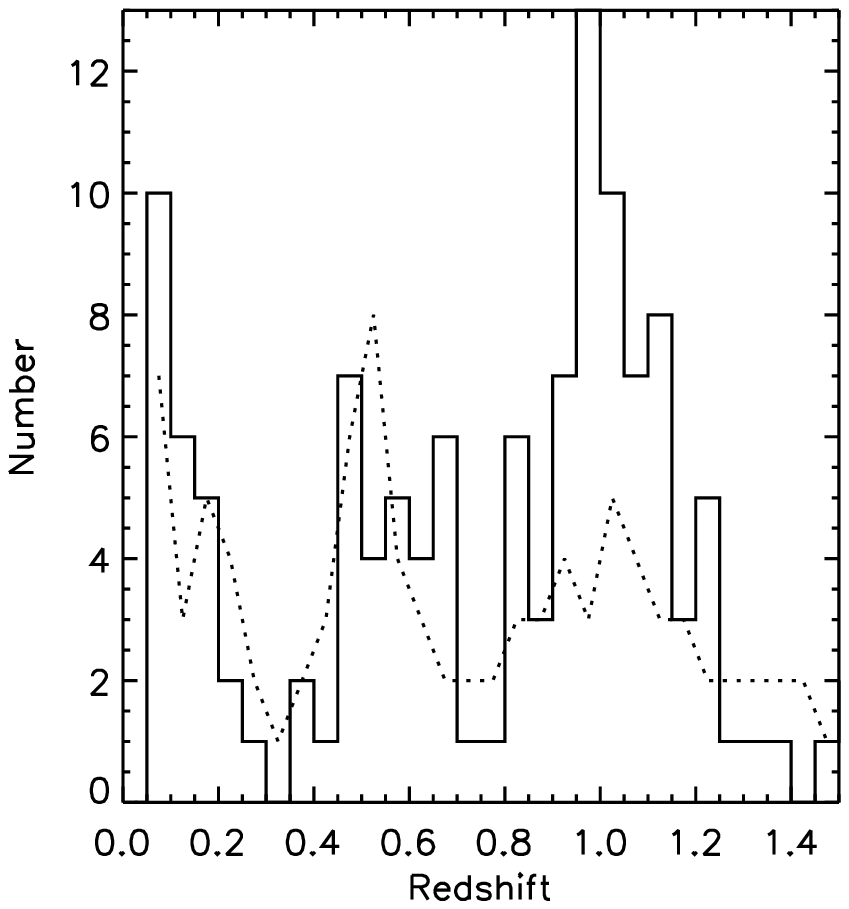}
\plotone{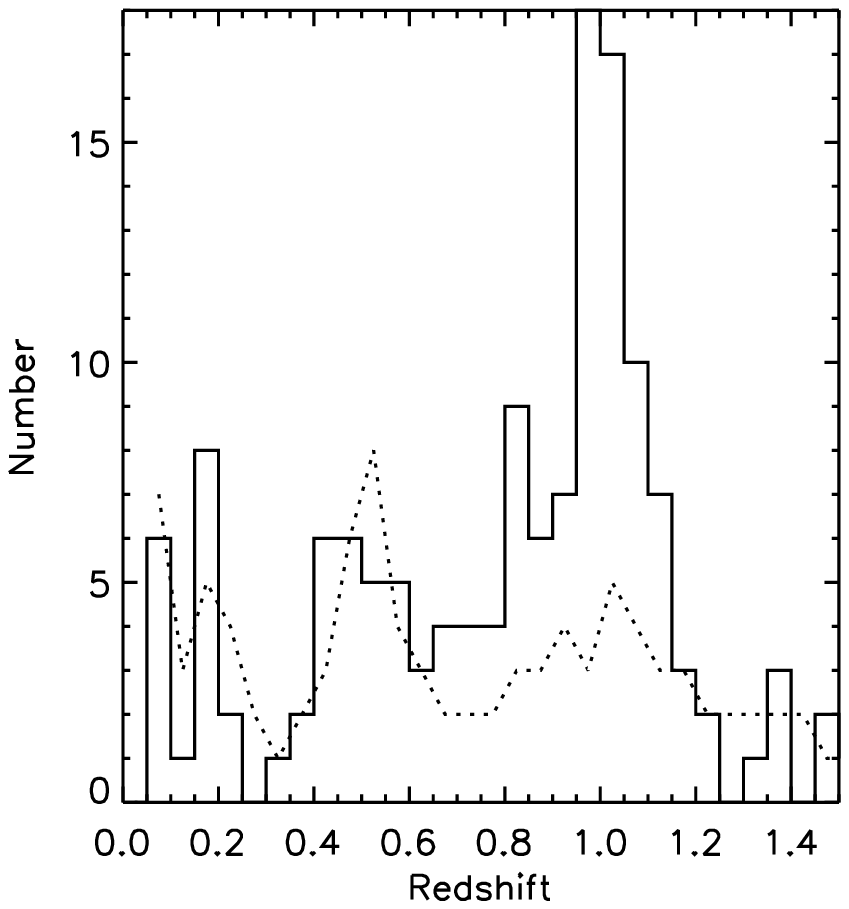}
\plotone{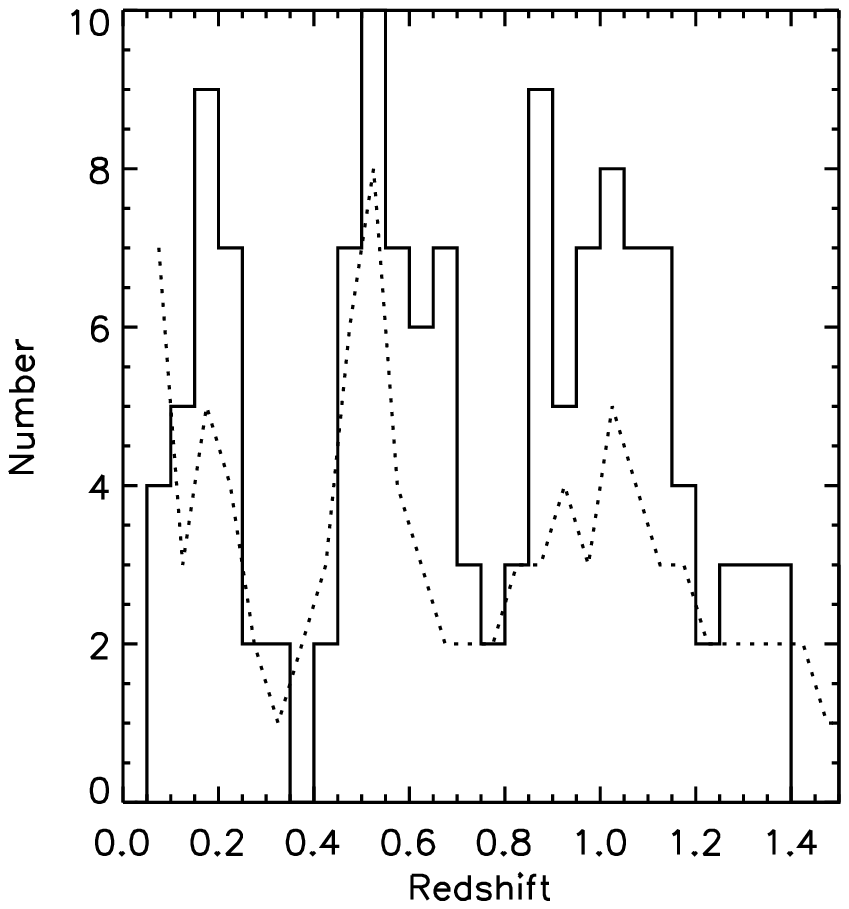}
\caption[Redshift distribution]{Redshift distributions of the three
  candidate clusters.  The solid histogram shows the redshifts for all
  galaxies within $\sim 375$\,Kpc of the cluster centers.  The dotted
  line shows the average redshift distribution of the field galaxy
  population measured in regions with the same area.  These
  distributions are based on photometric redshifts. }
\label{fig:zdist}
\epsscale{1}
\end{figure}

\begin{figure}
\epsscale{0.4}
\plotone{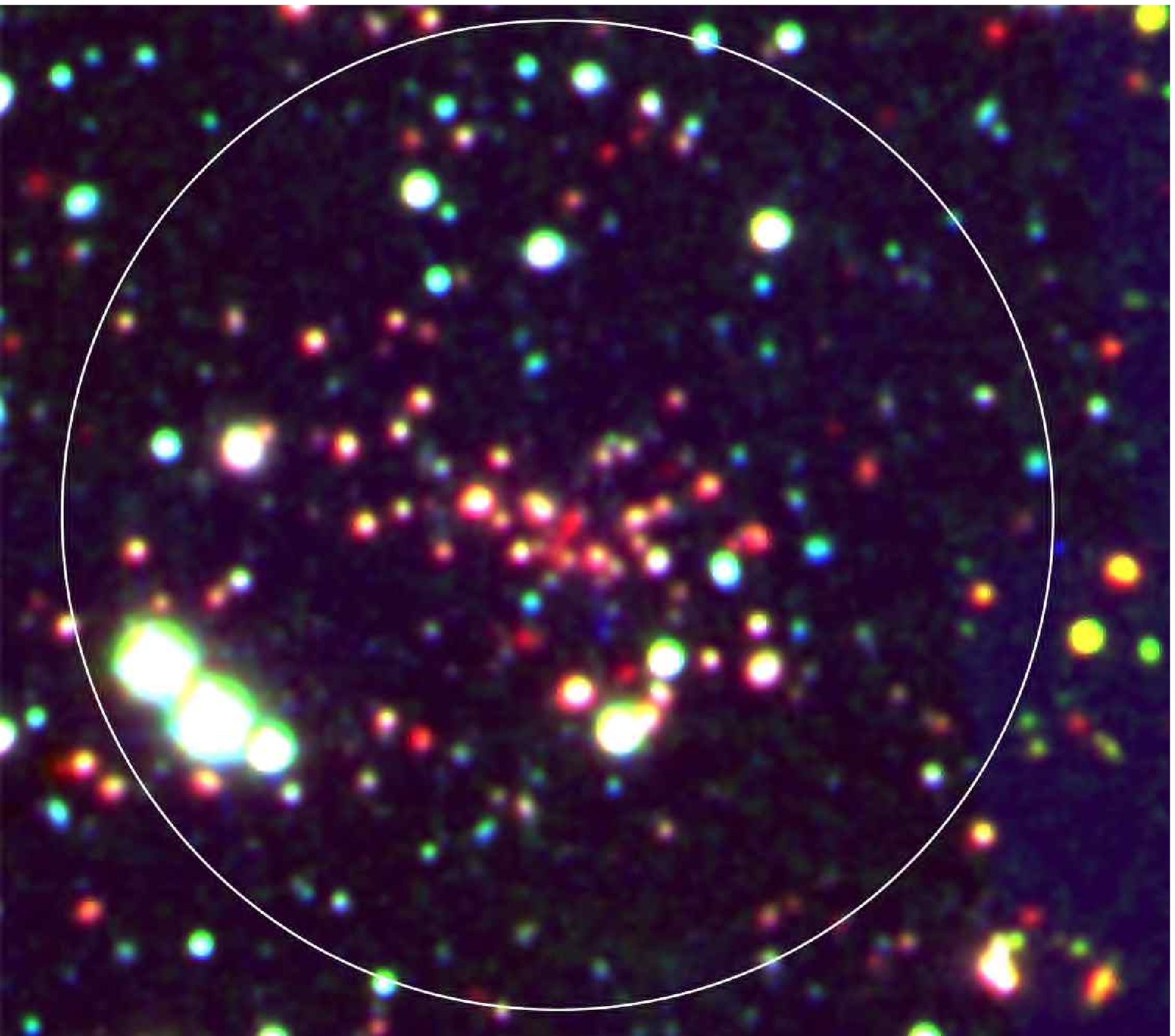}
\plotone{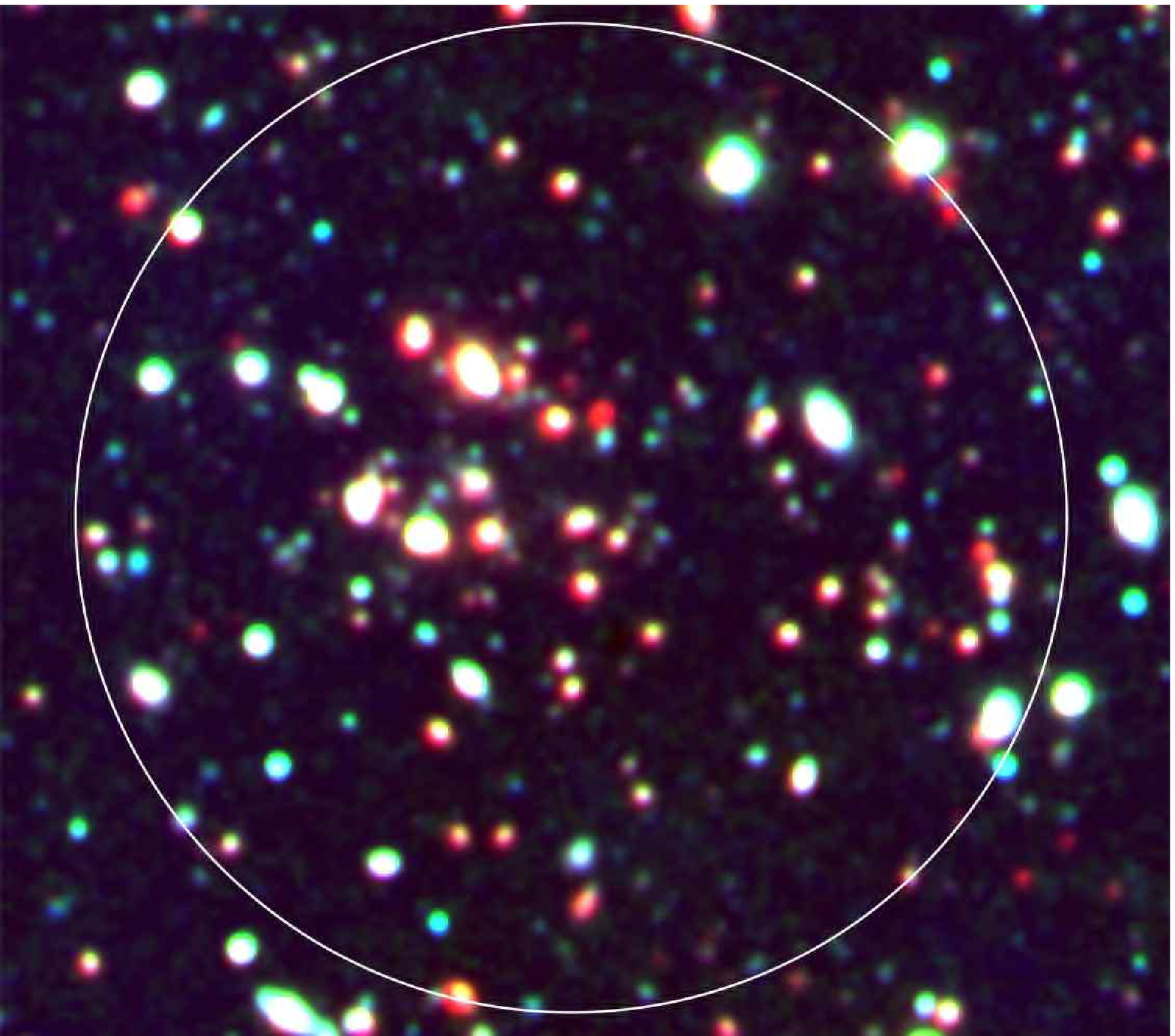}
\plotone{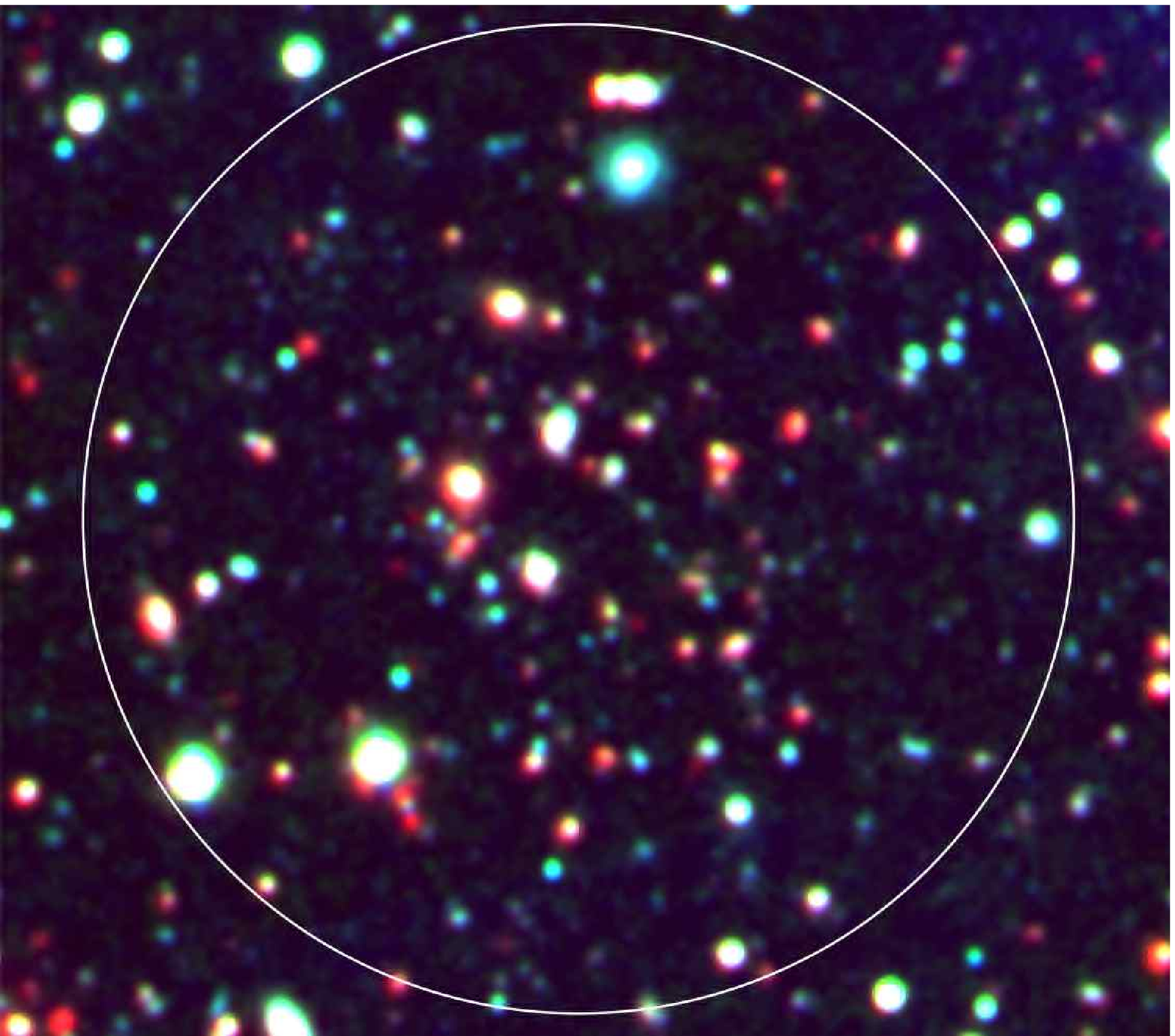}
%\plotone{acs_1.ps}
\caption[Cluster 1 ]{Color composite images of candidate clusters from
  ACS F814 (blue), MMT $z^\prime$ (green), and IRAC $3.6\,\micron$
  (red) data all smoothed to the resolution of the IRAC data.  The
  circle shows the 0.017\degr (500Kpc) radius used to count the number
  of galaxies on the red sequence.  }
\label{fig:clus1}
\epsscale{1}
\end{figure}

The first step in the search is to combine the spatial information
from the 2-D images with photometric redshifts to visually identify
clusters of galaxies.  We do this by plotting the locations of all
galaxies in a certain photometric redshift range on both the optical
and IR images to look for clusterings.  We step through redshift space
in overlapping z=0.2 bins examining each for clusterings.  The result
of this search is 14 cluster candidates at $0.8 < z < 1.4$ with
average areal densities of 10.3 galaxies with similar redshifts per
square arcminute.  Next, for each candidate cluster, we examine the
redshift distributions of all galaxies within 0.013\degr of the center
($\sim 375$ Kpc at these redshifts).  We choose this radius, at around
one third of the virial radius, as the size at which the clustering
signal appears to be strongest.  The candidate cluster redshift
distributions are compared to the average redshift distributions of 50
regions of the same area randomly distributed across the field.  After
this comparison, three excellent candidate clusters remain with peaks
in their redshift distributions which are greater than $2\sigma$ above
that of the comparison fields (see Figure \ref{fig:zdist} \& Table
\ref{tab:clusterchar}).  In addition the three best candidate clusters
show clear over-densities in both the ACS and IRAC images (see Figure
\ref{fig:clus1}).  While we cannot guarantee that this search is
complete without extensive spectroscopy, we are confident that we have
found the more massive clusters at redshift one.

In a shallow, large-area IRAC survey \citet{eisenhardt2008} finds
roughly two high-z clusters in an area equivalent to the IRAC dark
field.  The DEEP2 survey finds seven spectroscopically confirmed groups
and clusters with $0.75 < z < 1.03$ and an upper mass limit of
$1\times10^{13} \msun$ in a similar area to our survey
\citep{fang2007}.  Our finding of three candidate clusters is in agreement
with the number of clusters in these other surveys.

Two of the candidate clusters have centers within $1.8\arcmin$ of each
other.  If clusters were randomly distributed, from Monte Carlo
simulations we would expect to find a close pair like this one 2-50\%
of the time depending on the areal density (here using the IRAC
shallow survey and DEEP2 respectively).  However, from both
hierarchical simulations and cluster surveys we do not expect clusters
to be randomly distributed, instead clusters are connected by
filaments and are highly correlated even at redshift
one\citep{brodwin2007}.  If our two clusters are truly close in all
three dimensions then they must lie along a filament.  Our photometric
redshifts do not allow for the precision needed to know if they indeed
are 3-D neighbors.

\subsection{X-ray Properties}
\label{xray}

\begin{figure}
\epsscale{0.5}
\plotone{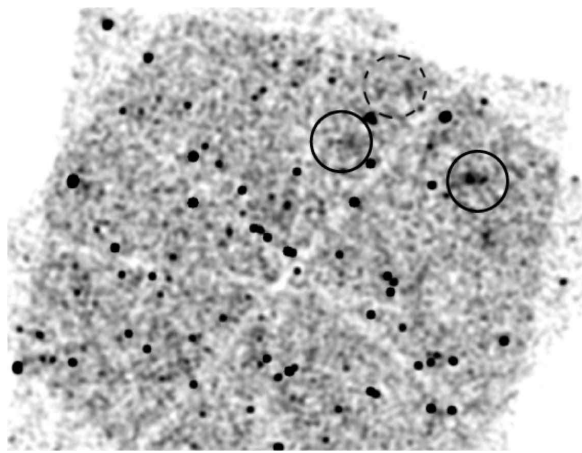}
\caption[chandra]{Chandra ACIS-I 0.3 - 7kev image of extended emission
  in the field.  The two detected clusters are encircled with a 500Kpc
  radius circle.  The third, undetected cluster location is shown with
  a dashed line circle.  The detected cluster on the right has a
  higher count rate and therefore a higher mass.  Specifically the
  extended emission is associated with the center of the cluster and
  not with the stars in the bottom left of Figure \ref{fig:clus1}.The
  extended emission just below the cluster on the right corresponds to
  a cluster at z$_{phot}$=0.25, and is not discussed in this paper. }
\label{fig:chandra}
\epsscale{1}
\end{figure}

Two of the three candidate clusters show extended emission in our
100ks Chandra ACIS-I image.  This both confirms that they are bound
clusters because at these detected luminosities they can only be
clusters, and allows us to determine their masses (see Figure
\ref{fig:chandra}).  We are also able to place a limit on the mass of
the third, Chandra undetected, cluster candidate.

X-ray luminosities are obtained by doing aperture photometry on the
merged ACIS-I image with the CIAO task {\it dmextract}.  Aperture
sizes were chosen to be 0.9\arcmin (0.4Mpc) which fully encompasses
all of the excess flux coming from the clusters but avoids neighboring
point sources on the Chandra image.  Point sources prevent us from
using background regions directly adjacent to the clusters, so instead
we use the average of 20 background regions taken from empty regions
all over the frame.  Net counts are 172 $\pm 31$ and 72 $\pm 29$
(Poisson errors) for the clusters respectively.  It is possible that
the brighter cluster (on the right in Figure \ref{fig:chandra}) is
contaminated by a point source which is effecting this count rate,
however with such low number of photons we have not attempted to
separate this possible point source from the cluster and background.

Taking the count rate from the image, we use the Portable Interactive
Multi-Mission Simulator (PIMMS v 3.9d) to derive a flux assuming a
thermal bremsstrahlung model with $4\times10^{20}$cm$^{-2}$ of Galactic
NH (measured for the location of this field), and a temperature of
5KeV.  This temperature is chosen randomly and is calculated more
accurately in the following iterative process.  Once we have
calculated a flux and luminosity of the clusters, we use a combination
of the \citet{maughan2006} and \citet{vikhlinin2002} $L_x - T $
relation based on Chandra and XMM data on 22 clusters with $z > 0.4$
to derive a more realistic temperature.  We then go back and use this
new temperature to re-calculate the X-ray flux and luminosity.  The
two clusters with detections have luminosities of $3.6\pm 0.6\times
10^{43}$ and $1.6 \pm0.7 \times 10^{43}$ erg/s in the 0.5-2 Kev band.

\subsection{Mass}
\label{mass}

To derive the mass from the X-ray luminosity for the two detected
clusters, we use the \citet{maughan2007} $L_x - M_{500}$ relation
based on 34 clusters with $0.5<z<1.3$.  Derived masses are $M_{500}=
6.2 \pm 1.4 \times 10^{13}$ and $3.6 \pm 1.4 \times 10^{13}$ \msun.
Since the second cluster is a very low count detection, we take this
mass to be the upper limit to the mass of the third, Chandra
undetected, cluster.  Quoted errors only represent
error in the measurements, including uncertainties in redshift, but do
not take into account error in the models and should therefore be
taken as optimistic.

There is some evidence that using X-ray luminosities to estimate mass
is unreliable at high redshift in the sense that higher redshift
clusters have lower x-ray luminosities for a given mass
\citep{lubin2004, fang2007}.  The DEEP2 survey of the aforementioned
seven clusters at $0.75< z< 1.0 $ detect none of their clusters in a
200ks Chandra image \citep{fang2007}.  A possible reason for this is
that clusters are not virialized at higher redshifts, an assumption
which is necessary to use the cluster hot gas to measure mass.
Alternatively, \citet{andreon2008} have proposed that there are no
underluminous X-ray clusters and that previous work has either not
properly measured cluster mass via velocity dispersion, or have not
correctly interpreted their X-ray observations.  If high-z clusters
are underluminous in X-rays then there are two relevant implications:
1) our high-z clusters have true masses which are higher than measured
from their Chandra luminosities, and 2) this leaves open the
possibility that the Chandra undetected cluster is a true cluster with
mass larger than the limit afforded by the current non-detection. The
resolution of this issue is beyond the scope of this paper.

As a check of cluster mass we have considered using the SDSS relation
of optical richness to the weak lensing mass of clusters at low
redshifts \citep{johnston2007}.  This relies on multiple assumptions
and relations such as that between the number of red sequence galaxies
inside of $1h^{-1}$Mpc with luminosities greater than $0.4L^*$ and
$r_{200}$, the radius at which the cluster is 200 times the critical
density of the universe \citep{hansen2007} .  Furthermore one has to
then use the derived value of $r_{200}$ with the weak-lensing relation
to arrive at $M_{200}$.  To our knowledge none of these relations have
been tested at high redshifts and we therefore choose not to make this
calculation.

The measured mass of these clusters is consistent with the masses of
the, to date, roughly 35 published groups and clusters with confirmed
redshifts above 0.9 \citep{stanford1997, ebeling2001, stanford2001,
  stanford2002,eblanton2003, rosati2004, margoniner2005, mullis2005,
  siemiginowska2005,elston2006, stanford2006, fang2007,
  finoguenov2007, eisenhardt2008}.  About two thirds of these clusters
are detected in multi-wavelength surveys such as DEEP2, COSMOS, and
the IRAC shallow survey and have an average mass of
$3.8\times10^{13}\msun$.  The rest are detected in serendipitous X-ray
imaging or are targeted for their location around radio galaxies.
These other clusters all have higher masses ($>1E14\msun$) which is
consistent with it being easier to detect high-mass clusters in the
X-ray and high mass clusters around radio galaxies.  Overall we expect
from hierarchical formation that clusters at $z=1$ should be less
massive than clusters at $z=0$.
%---------------------------------------------------------------------------------

\section{Color Magnitude Diagrams}
\label{cmds}

\begin{figure}
\epsscale{0.4}
\plotone{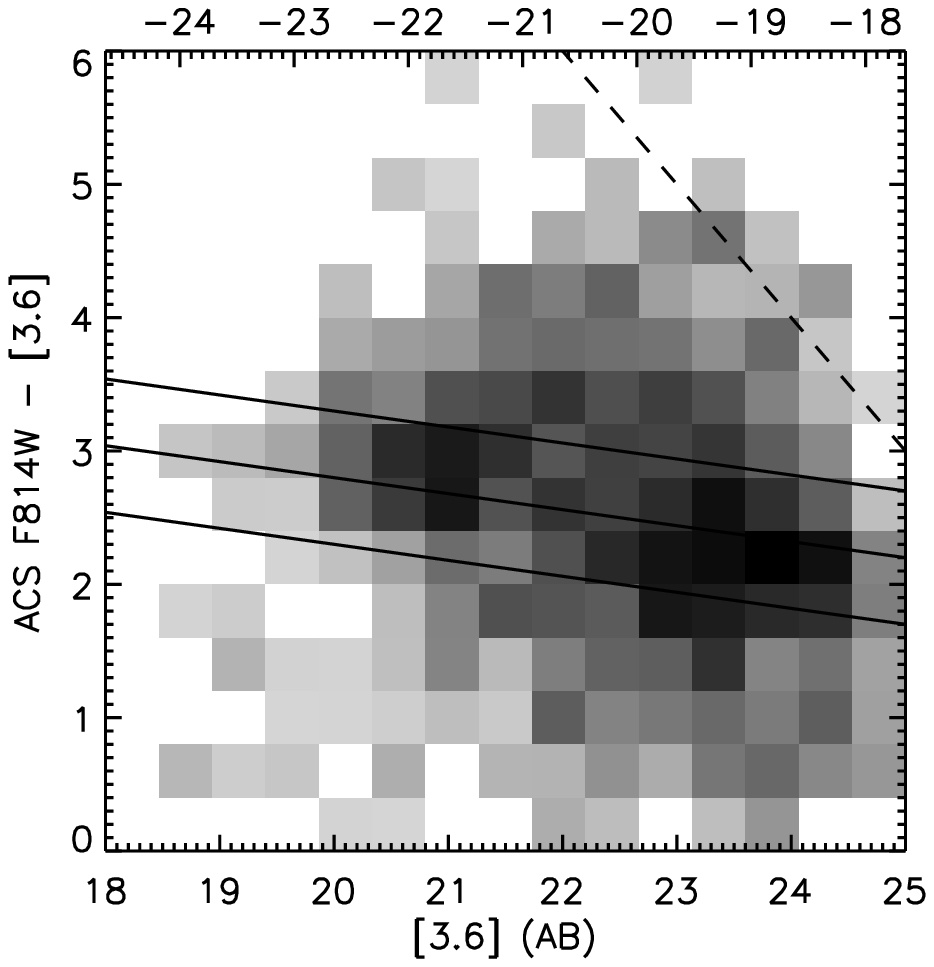}
\plotone{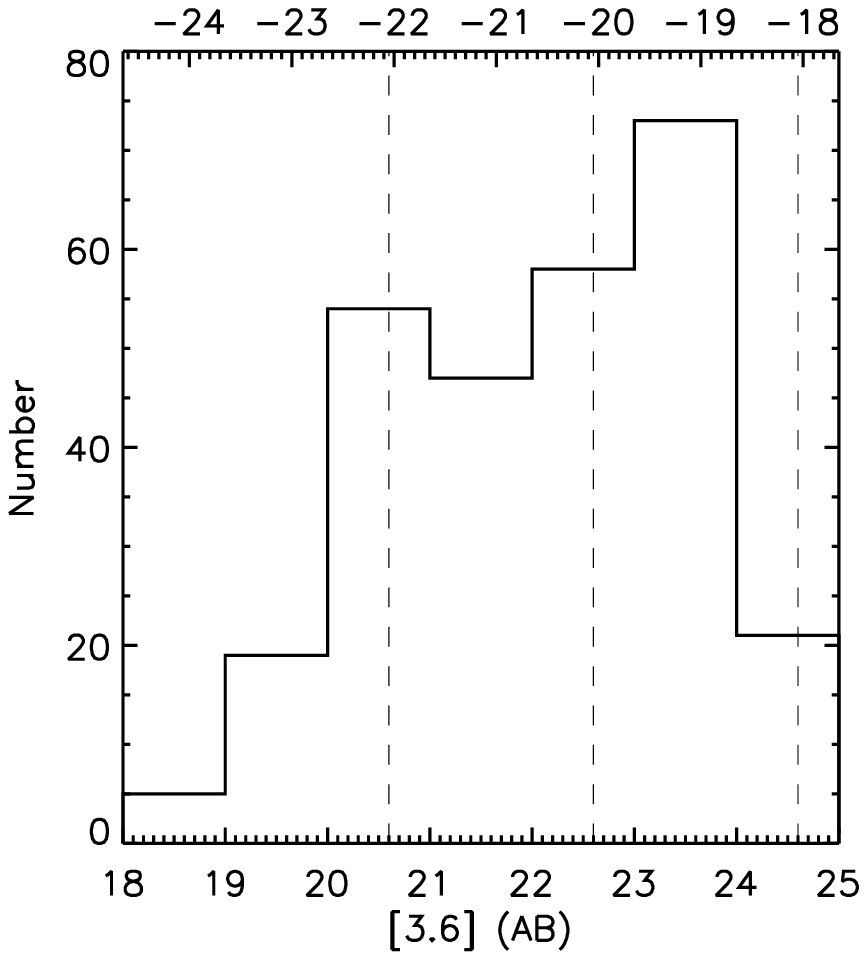}
\caption[cmd]{Left: Color magnitude diagram of all 711 galaxies within
  500\,Kpc of the centers of the three clusters at z=1 from the IRAC
  dark field. Greyscale indicates the density of galaxies within each
  bin.  Lines represent the RCS $\pm 0.5$mag.  The top scale is
  absolute rest frame K-band magnitude.  The dashed line shows the
  $5\sigma$ detection limit of the ACS data which does not come close
  to our fitted RCS. The CMD is plotted only to magnitudes well
  brighter than the measured 95\% completeness limit of the IRAC
  data. Right: Distribution of galaxies along the red sequence for the
  composite cluster at z=1. Dashed lines delineate the faint and
  bright magnitude bins used in the analysis.}
\label{fig:cmd}
\epsscale{1}
\end{figure}

Most galaxies in clusters are ellipticals that have approximately the
same red color regardless of magnitude and therefore form a red
cluster sequence (RCS) in a color magnitude diagram (CMD), as long as
the color is chosen to span the 4000 \AA\ break.  Past studies with
HST have shown that the RCS is in place at least by redshift one, if
not before \citep{blakeslee2003,vandokkum2001}.  We choose to make
CMDs of the dark field clusters with ACS F814W and Spitzer
$3.6\,\micron$.  At $z\sim1$ this corresponds to rest-frame B-K which
does span the Balmer break.  We generate a composite color magnitude
diagram for all three clusters including all galaxies within
0.017\degr ($\sim 500$ Kpc) of the candidate cluster centers (Figure
\ref{fig:cmd}).  The upper axis is plotted as absolute K-band
magnitude in order to compare to a low redshift sample.  Because
$3.6\,\micron$ at $z=1$ corresponds to K-band at $z\simeq0.1$, we do not
have to apply a K-correction to the data, which avoids a large set of
uncertainties.  We do make a correction for the redshifting of the
bandpass and a correction for the luminosity evolution of galaxies
over time taken from the stellar evolution models of
\citet{bruzual2003}.  The color of RCS is consistent with the
predicted colors from simple stellar populations with a single burst
of star formation above z=3 and a Salpeter initial mass function
\citep{bruzual2003}.

Without benefit of spectroscopy we determine membership based on
location in the CMD.  Pinpointing the location of the RCS becomes
easier when using a composite CMD from three clusters.  To determine
the location of the RCS, the bright end of the composite red sequence
($[3.6] < 22.5$) is fit with a biweight function \citep{beers1990}
where the slope is fixed to -0.12 (to match the low redshift slope
below).  The faint end is left out of the fit because of the larger
amount of contamination by field galaxies.  Members are taken to be
those galaxies which lie within $\pm 0.5$ mag of the RCS. We determine
the significance of the existence of a red sequence in our CMD by
doing a monte carlo calculation with 100 realizations of the CMDs of
randomly selected galaxies in the dataset.  From this calculation we
find that the number of galaxies on the true RCS is significant
compared to random galaxies at the $6.7\sigma$ level.  The histogram
of member galaxies in the composite cluster is shown in Figure
\ref{fig:cmd}.

Although the RCS is mainly composed of cluster galaxies at a
particular redshift, there will be some fore- and back-ground galaxies
that will contaminate the membership count. We statistically subtract
fore-and back-ground galaxies within the measured RCS.  The number of
contaminating galaxies is calculated by counting the average number of
galaxies from 50 random regions of the same size as the clusters which
have colors consistent with the measured RCS.  The average number of
contaminating galaxies within a cluster area (0.017\degr) and their
standard deviation are $40.3\pm7.3$ and $16.9\pm8.0$ respectively for
the magnitude bins used below ( $-20 < M_k < -18$ and $-22 < M_k <
-20$).  After this subtraction the number of member galaxies remaining
in each cluster is a few to none in the faint bin, and about twenty
each in the bright bin.

\begin{figure}
\epsscale{0.4}
\plotone{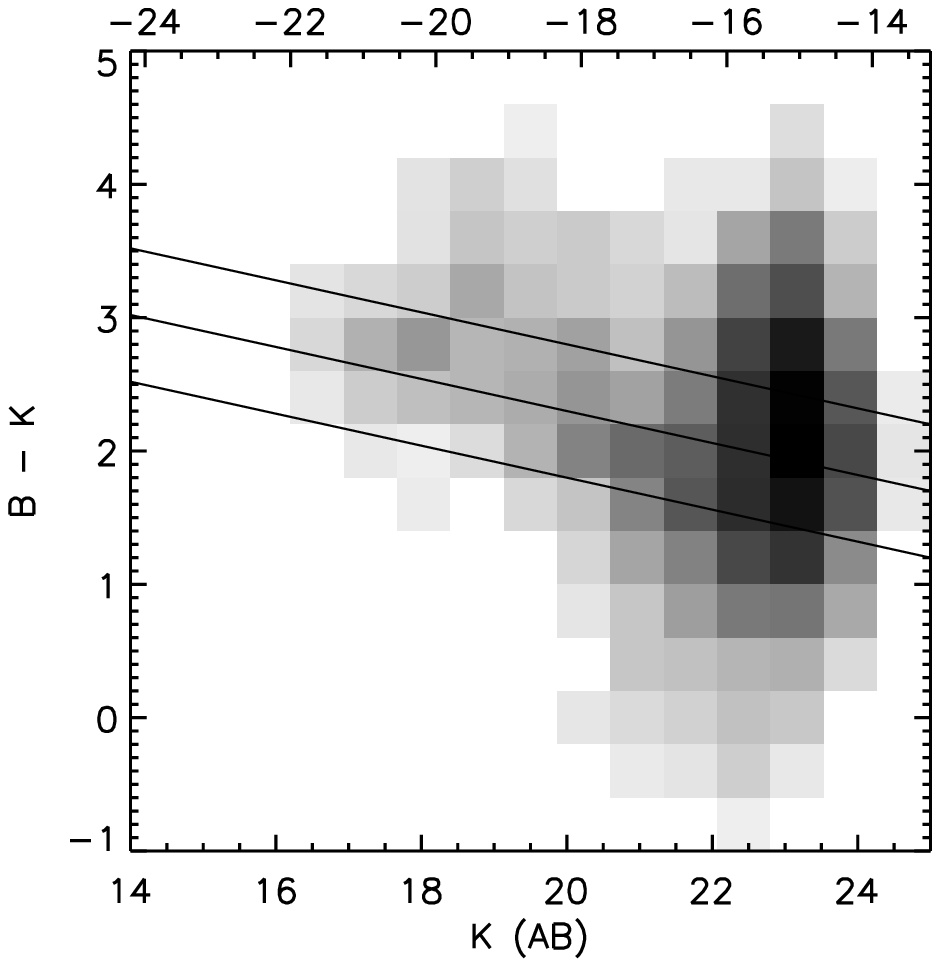}
\plotone{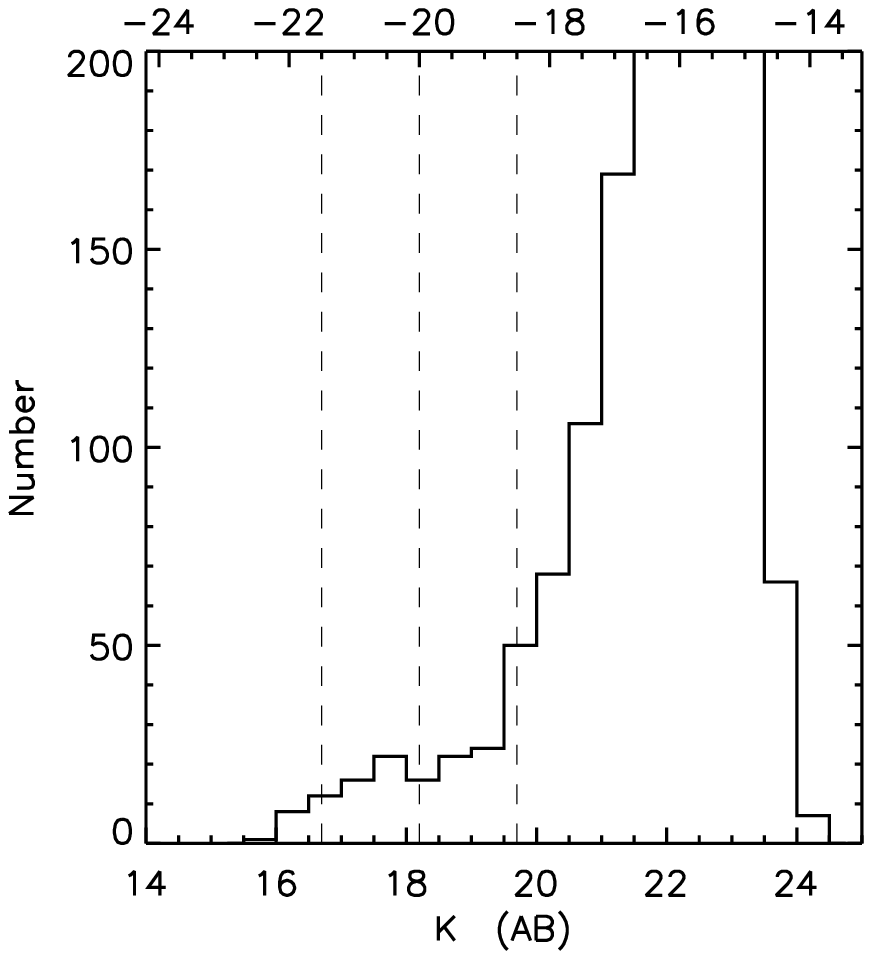}
\caption[cosmoscmd]{Color magnitude diagram and K-band distribution
  for the composite COSMOS cluster at z=0.1.  Lines have the same
  meaning as Figure \ref{fig:cmd}.  For clarity this CMD only includes
  those galaxies with photometric redshifts in the range of the
  clusters, however the actual measurement is made in the exact same
  way as for the clusters at $z=1$ which is to use the RCS minus a
  statistical background for membership information.}
\label{fig:cosmoscmd}
\epsscale{1}
\end{figure}

We create CMDs for a comparison sample of low redshift clusters using
the same process on publically available B- and K- band data from the
Cosmic Evolution Survey \citep[COSMOS;][]{capak2007}.  From COSMOS we
use four X-ray confirmed clusters with average $0.09 < z < 0.12$
\citep[][Id's 42, 58, 113, 140]{finoguenov2007}.  These clusters have
an average mass of $M_{500} = 1.1\times 10^{13} \msun$ as determined
from the X-ray temperature.  Figure \ref{fig:cosmoscmd} shows both the
CMD of the composite COSMOS data from four clusters and the K-band
distribution of RCS member galaxies including a statistical correction
for fore- and back-ground galaxies taken from neighboring COSMOS
regions.

This comparison sample is not ideal since these low redshift clusters
are roughly the same mass as the high-z clusters, and we expect
clusters to hierarchically gain mass over time through the infall of
other groups and clusters.  Unfortunately, to our knowledge, there is
no sufficiently deep near-IR imaging on local $\sim 10^\{14\}$\msun
clusters.  While it is possible that more massive clusters will have
fewer faint red galaxies than a less massive cluster due to merging,
we do not expect this to be a large effect.  Deep, wide-field, near-IR
imaging of local clusters is required to study this effect in detail.

The CMD's from the two samples have clear differences.  Despite our
ultradeep data, the distribution of the $z=0.1$ galaxies extends to
fainter magnitudes than that for the clusters at $z=1$ (thanks to the
approximately four magnitudes of surface brightness extinction).  We
plot the deeper low redshift data to $M_K =-14$, whereas for the high
redshift dark field clusters we can only plot a smaller range of
magnitudes extending to $M_k=-18$.  Everything fainter than that is
below our $5\sigma$ detection threshold, so we are unable to compare
that magnitude range with the lower redshift clusters.
%---------------------------------------------------------------------------------

\section{Results}
\label{results}

We investigate the evolution of the fraction of faint galaxies on the
red sequence from redshift one to the present.  To make the comparison
between the two different redshift samples, we choose a rest-frame
K-band absolute magnitude range which is both brighter than the $5
\sigma$ detection threshold for both datasets, and not too far towards
the bright end of the luminosity function that extremely small number
statistics would effect the measurement.  We divide the resulting
magnitude range into two bins which are two magnitudes wide; $-20 <
M_k < -18$ (faint bin), and $-22 < M_k < -20$ (bright bin).

\begin{figure}
\epsscale{0.5}
\plotone{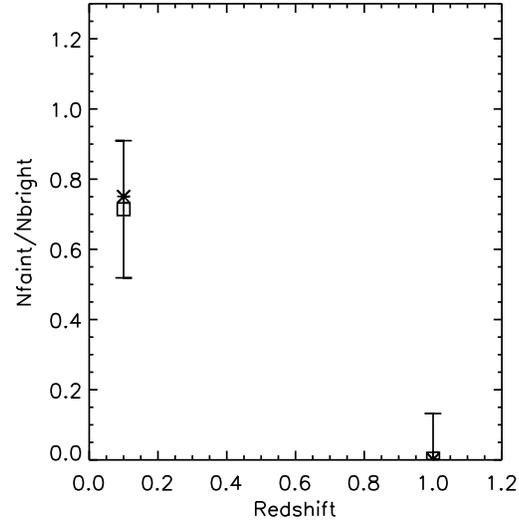}
\caption[ratio]{Ratio of faint-to-bright galaxies on the red sequence
  in both the samples at high and low z.  Squares represent background
  subtracted numbers of member galaxies on the red sequence measured
  inside of a 500 KPC radius.  Stars show the same measurement for a
  750 Kpc radius.  Error bars are only shown on the 500 Kpc points and
  represent error in the background measurement.  }
\label{fig:ratio}
\epsscale{1}
\end{figure}

The quantity we want to measure is the amount of faint galaxies on the
RCS at $z=1$ compared to the present epoch as an evolutionary measure
of the build-up of the red sequence.  This is calculated as the ratio
of faint RCS galaxies to bright RCS galaxies.  This faint-to-bright
ratio is shown explicitly for both cluster samples in Figure
\ref{fig:ratio} and can also be seen by comparing Figures
\ref{fig:cmd} \& \ref{fig:cosmoscmd}.  We find that the relative
number of faint galaxies on average in clusters at redshift one is
less than in the average cluster at redshift 0.1 at the $3\sigma$
level.

The major sources of error in this work are the background measurement
of field galaxies, the area over which cluster membership is
determined, the possibility that the third cluster is a chance
projection, and the size of the magnitude bins. Recall that
completeness in the IRAC data is not a limiting factor to this
measurement because we use ACS source locations and shapes as a prior
for doing IRAC photometry.  To quantify the error originating from the
field galaxy subtraction, error bars on Figure \ref{fig:ratio} show
one standard deviation on the background measurement propagated to the
ratio of faint to bright number of galaxies.

To estimate the effect of changing the area over which cluster members
are counted we re-calculate the faint-to-bright ratio using a radius
which is 50\% larger than the original radius (2.25 times the area).
The original radius was chosen as a compromise between the relative
physical sizes of the implied area around the low redshift and high
redshift clusters.  For reference the original 500\,Kpc radius circle
is shown on the IRAC images in Figure \ref{fig:clus1}.  The new,
larger radius, ratios are shown as stars in Figure \ref{fig:ratio}.
The effect of changing radius is negligible.

We examine the possibility that the third cluster is a chance
projection, not a cluster, and therefore contaminating this
measurement.  Spectroscopy is the best way to confirm that it is a
cluster.  Without spectroscopy, we re-calculate the ratios without the
third cluster, using just the two X-ray detected clusters.  The same
trend of the bright galaxies far outnumbering the faint galaxies on
the red sequence at z=1 is recovered, albeit with a larger noise
contribution.  Additionally, we test that any two clusters are
dominating the signal by removing one cluster from the measurement in
turn. When we recalculate ratios each time the significance of the
signal goes down.  From this we conclude that all three clusters are
contributing to the signal.

Lastly, we consider the possibility that the four magnitude range used
is too large in that it goes too close to the very uncertain bright
end of the luminosity function of the z=0.1 clusters, and too far into
a regime where photometry is confused at the faint end of the
luminosity function in the z=1.0 clusters.  To test our results, we
re-calculate the ratios with 1.5 magnitude wide bins centered at $M_K
=-20$.  We find, again, the same trend of faint galaxies disappearing
from the red sequence population at high redshift.

There is a clear deficit of faint galaxies on the red sequence at z=1
compared to the current epoch when taking into account possible
sources of error.

%---------------------------------------------------------------------------------

\section{ Discussion}
\label{discuss}

We find that there are fewer faint red cluster galaxies at high
redshift than low redshift in comparison to the number of bright red
galaxies.  There are two possible explanations for this.  There could be
overall less faint cluster galaxies at z=1 at all colors than the
present epoch, or the faint end of the red sequence is not yet in
place, and instead those galaxies which will fill the faint end at
z=0, are still forming stars at higher redshifts.

The first explanation, that there are overall, at all colors, fewer
faint galaxies at z=1 than at z=0, goes directly against the theory of
hierarchical formation.  We expect that clusters at higher redshifts
will have more faint galaxies than today, and that over time the faint
galaxies will merge into brighter, more massive galaxies.  This will
have the effect of clusters at higher redshifts having steeper faint
end slopes of the luminosity function than today's flatter slopes, or
exactly the opposite trend of what would be suggested if overall there
were less faint galaxies at $z=1$.  While we do not think this is a
likely solution, it is interesting that in a simulation of cluster
luminosity functions, \citet{khochfar2007} find a slight trend that
the faint end slope of clusters does steepen from redshift 1 to 0,
whereas overall they find the hierarchical formation trend of
flattening slope from z=6 to 0.  This is probably due to noise in
their calculation.  It would be nice to know how the overall faint end
slope of clusters is evolving from z=1 to 0, however we do not have a
large enough sample to attack this problem.

It is more likely that the deficit of faint red galaxies at high z is
due to the fact that those faint galaxies are still forming stars, and
so have bluer colors at those redshifts.  This implies that a) bright,
massive galaxies have already shut off their star formation in
clusters by z=1, and b) faint, less massive galaxies are still forming
stars in clusters at z=1.  We know that clusters
exhibit red sequences at higher z, so it is not new that massive
galaxies have already undergone some process, excited by the cluster
environment, which keeps them from forming stars, and lands them on
the red sequence.  This is further evidence for the popular theory of
'downsizing', in which the more massive galaxies evolve first.  However, all
galaxies in clusters do not follow the same evolutionary processes,
instead evolution from the blue cloud to the red sequence seems to be
mass dependent.  Whatever process stops star formation in clusters
appears has not yet happened at z=1 to the less massive galaxies.
Those galaxies are still being allowed to form enough stars to stay
off of the red sequence.  The red population of clusters is not yet
fully in place by z=1.

We do not compare the specific values of faint-to-bright ratios of this
work with those in the literature as different wavelengths, areas, and
definitions of faint and bright are used.  However the trend for a
deficit of faint galaxies at high redshifts is in agreement with the
works of \citet{delucia2007, gilbank2008, stott2007} and
\citet{koyama2007}, but in contrast to \citet{andreon2007}.  The cause
of the difference is unclear, but we note that we use a much redder,
wider filter set at high redshifts (rest-frame B-K) than
\citet{andreon2007} (rest-frame u-g).  We also compare with the same
rest-frame color at low redshift without applying a k-correction.

Future work in this field requires a large enough sample of confirmed
clusters at high-z with consistent observations well below $M^*$ to be
able to split the sample on cluster properties.  There are hints that
the faint-to-bright ratios are cluster mass or richness dependent, but
the literature shows contradictory trends for these effects likely due
to the small samples used to date \citep{gilbank2008, delucia2007}.
With this evidence for ongoing star formation in clusters at z=1, in
paper two of this series we will examine the star formation rates and
morphologies of the cluster galaxies using deep SPITZER MIPS
$24\,\micron$ and HST ACS F814W data. Additionally as the community continues to
build a larger sample of high redshift clusters we will be able to
study their properties, in particular their suitability for dark
energy number count surveys \citep{wang2004}.
%---------------------------------------------------------------------------------

\acknowledgments

We acknowledge E. Rykoff for help with the mass measurement.  We thank
the anonymous referee for useful suggestions on the manuscript. This
research has made use of data from the Two Micron All Sky Survey,
which is a joint project of the University of Massachusetts and the
Infrared Processing and Analysis Center/California Institute of
Technology, funded by the National Aeronautics and Space
Administration and the National Science Foundation.  This work was
based on observations obtained with the Hale Telescope, Palomar
Observatory as part of a continuing collaboration between the
California Institute of Technology, NASA/JPL, and Cornell University,
the Spitzer Space Telescope, which is operated by the Jet Propulsion
Laboratory, California Institute of Technology under a contract with
NASA, the MMT Observatory, a joint facility of the Smithsonian
Institution and the University of Arizona, and the NASA/ESA Hubble
Space Telescope, obtained at the Space Telescope Science Institute,
which is operated by the Association of Universities for Research in
Astronomy, Inc., under NASA contract NAS 5-26555. These observations
are associated with program \#10521.  Support for program \#10521 was
provided by NASA through a grant from the Space Telescope Science
Institute, which is operated by the Association of Universities for
Research in Astronomy, Inc., under NASA contract NAS 5-26555.

{\it Facilities:} \facility{Palomar (LFC, WIRC)}, \facility{MMT
  (Megacam)}, \facility{HST (ACS)}, \facility{Spitzer (IRAC, MIPS)},
\facility{Akari}, \facility{CXO (ACIS)}

%------------------------------------------------------------------------
%\clearpage
%\bibliography{ms.bbl}  

%\bibliography{jkrick}  
%------------------------------------------------------------------------

% Table generated by Excel2LaTeX from sheet 'Sheet1'
\begin{deluxetable}{ccccccc}
\tablewidth{0pc}
\tablecolumns{7}
\tablecaption{Cluster Characteristics \label{tab:clusterchar}}

\tablehead{
\colhead{Cluster} &         
\colhead{ra} &        
\colhead{dec} &      
\colhead{$N_{gals}$} & 
\colhead{z$_{peak}\tablenotemark{a}$} & 
\colhead{L$_{x}$ (0.5-2.0 Kev)} &    
\colhead{M$_{500}$}
\\
\colhead{ } &
\colhead{J2000 (deg)} &      
\colhead{J2000 (deg)} &  
\colhead{$r < 500 Kpc$} &
\colhead{ }            & 
\colhead{$1\times10^{43}$erg/s} & 
\colhead{$1\times10^{13}\msun$}
}

\startdata
         1 &  264.68160 &   69.04481 &        215 & $1.0\pm0.1$ & $3.6\pm0.6$ & $6.2\pm1.4$ \\

         2 &  264.89228 &   69.06851 &        255 & $1.0\pm0.1$ & $1.6\pm0.7$ & $3.6\pm1.4$ \\

         3 &  264.83102 &   69.09031 &        241 & $1.0\pm0.2$ & $\le1.6\pm0.7$ & $\le3.6\pm1.1$ \\

\enddata
\tablenotetext{a}{Redshift peak and one sigma uncertainty are measured from a gaussian fit to the redshift distribution.}
\end{deluxetable}  
%------------------------------------------------------------------------

\clearpage

%-----------------------------------------------------------------

%-----------------------------------------------------------------

%-----------------------------------------------------------------

%-----------------------------------------------------------------
%-----------------------------------------------------------------

%-----------------------------------------------------------------

\end{document}